\documentclass[ reprint, aps,]{revtex4-1}
\usepackage{amsfonts}
\usepackage{amssymb}
\usepackage{amsmath}
\usepackage{graphicx}
\usepackage{dcolumn}
\usepackage{bm}

\begin{document}

\title{New kind of asymmetric integration projection operators constructed
by entangled state representations and parity measurement}
\author{Wang Shuai$^{1, \dagger }$, Wang Zhi-Ping$^{2}$ and Zhang Jian-Dong$%
^{1}$}
\address{$^1$ School of Mathematics and Physics, Jiangsu University of Technology,
Changzhou 213001, P.R. China
\\$^{\dagger }$ Corresponding author: wshslxy@jsut.edu.cn}

\address{$^2$ Department of Fundamental Courses, Wuxi Institute of
Technology, Wuxi 214121, P.R. China}

\begin{abstract}
By means of the technique of integration within an ordered product of
operators and Dirac notation, we introduce a new kind of asymmetric
integration projection operators in entangled state representations. These
asymmetric projection operators are proved to be the Hermitian operator.
Then, we rigorously demonstrate that they correspond to a parity measurement
combined with a beam splitter when any two-mode quantum state passes through
such device. Therefore we obtain a new relation between a Hermitian operator
and the entangled state representation. As applications, we recover the
previous results of the parity measurement in quantum metrology by our
formalism.

\begin{description}
\item[Keywords] quantum mechanics, Dirac notation, IWOP, entangled state
representation
\end{description}
\end{abstract}

\maketitle

\section{Introduction}

The transformation theory of quantum representations and Dirac's ket-bra
operators play important roles not only in quantum mechanics, but also in
information optics and quantum optics. Especially, with the help of the
technique of integration within an ordered product of operators (IWOP) \cite%
{1}, one can introduce many known or new quantum unitary operators by
constructing asymmetric integration projection operators based on quantum
representations. In this way, quantum optical version of some classical
optical transformations such as optical Fresnel transformation, Hankel
transformation, fractional Fourier transformation, Wigner transformation,
wavelet transformation and Fresnel-hadmard combinatorial transformation have
been established \cite{2,3,4,5}. In addition, one can not only find some new
quantum unitary operators corresponding to the known optical
transformations, but also can reveal some new classical optical
transformations. For example, the two-mode squeezing operator holds a
natural expression in the entangled state representation \cite{6}. The
complex fraction Fourier transformation has also been introduced based on
the two mutually conjugate entangled state representations $\left\vert \eta
\right\rangle $ and $\left\vert \xi \right\rangle $ \cite{4}.

Besides the usual coordinate and momentum representations, as well as the
coherent state representation, entangled state representations are also
important quantum representations \cite{1}. The concept of quantum
entanglement is originated in the paper of Einstein, Podolsky and Rosen
(EPR) \cite{7}, arguing on the incompleteness of quantum mechanics.
According to the original idea of EPR, Fan and Klauder have introduced two
kinds of entangled state representations \cite{8}. One kind of the entangled
state representation $\left\vert \eta =\eta _{1}+i\eta _{2}\right\rangle $
is the common eigenvector of two particles' relative position $\hat{x}_{1}-%
\hat{x}_{2}$ and total momentum $\hat{p}_{1}+\hat{p}_{2}$, i.e., $\left(
\hat{x}_{1}-\hat{x}_{2}\right) \left\vert \eta \right\rangle =\sqrt{2}\eta
_{1}\left\vert \eta \right\rangle $, $\left( \hat{p}_{1}+\hat{p}_{2}\right)
\left\vert \eta \right\rangle =\sqrt{2}\eta _{2}\left\vert \eta
\right\rangle $. The explicit form of such common eigenvector is
\begin{equation}
\left\vert \eta \right\rangle =\exp \left[ -\frac{1}{2}\left\vert \eta
\right\vert ^{2}+\eta \hat{a}^{\dag }-\eta ^{\ast }\hat{b}^{\dag }+\hat{a}%
^{\dag }\hat{b}^{\dag }\right] \left\vert 0\right\rangle _{a}\left\vert
0\right\rangle _{b},  \label{1}
\end{equation}%
where $\hat{a}^{\dag }$ and $\hat{b}^{\dag }$ are two bosonic creation
operators, and $\left\vert 0\right\rangle _{a}\left\vert 0\right\rangle _{b}$
is a two-mode vacuum state. The entangled state $\left\vert \eta
\right\rangle $ is proved to be orthonormal and complete,%
\begin{equation}
\left\langle \eta ^{\prime }\right. \left\vert \eta \right\rangle =\pi
\delta \left( \eta ^{\prime }-\eta \right) \delta \left( \eta ^{\prime \ast
}-\eta ^{\ast }\right) \text{,}\int d^{2}\eta |\eta \rangle \langle \eta
|/\pi =\hat{1},  \label{2}
\end{equation}%
where $d^{2}\eta =d\eta _{1}d\eta _{2}$. On the other hand, its conjugate
state $\left\vert \xi =\xi _{1}+i\xi _{2}\right\rangle $ is the common
eigenvector of two particles' center-of-mass coordinate $\hat{x}_{1}+\hat{x}%
_{2}$ and relative momentum $\hat{p}_{1}-\hat{p}_{2}$, i.e., $\left( \hat{x}%
_{1}+\hat{x}_{2}\right) \left\vert \xi \right\rangle =\sqrt{2}\xi
_{1}\left\vert \xi \right\rangle $, $\left( \hat{p}_{1}-\hat{p}_{2}\right)
\left\vert \xi \right\rangle =\sqrt{2}\xi _{2}\left\vert \xi \right\rangle $%
. The explicit form of this common eigenvector is

\begin{equation}
\left\vert \xi \right\rangle =\exp \left[ -\frac{1}{2}\left\vert \xi
\right\vert ^{2}+\xi \hat{a}^{\dag }+\xi ^{\ast }\hat{b}^{\dag }-\hat{a}%
^{\dag }\hat{b}^{\dag }\right] \left\vert 0\right\rangle _{a}\left\vert
0\right\rangle _{b},  \label{3}
\end{equation}%
which also holds orthonormal relation and completeness relation,

\begin{equation}
\left\langle \xi ^{\prime }\right. \left\vert \xi \right\rangle =\pi \delta
\left( \xi ^{\prime }-\xi \right) \delta \left( \xi ^{\prime \ast }-\xi
^{\ast }\right) \text{,}\int d^{2}\xi |\xi \rangle \langle \xi |/\pi =\hat{1}%
,  \label{4}
\end{equation}%
where $d^{2}\xi =d\xi _{1}d\xi _{2}$. By means of IWOP, these two mutually
conjugate entangled state representations $\left\vert \eta \right\rangle $
and $\left\vert \xi \right\rangle $ are very useful in information optics
and quantum optics \cite{1}. Nowadays, entangled state representations have
been widely used in the squeezed state theory, the Wigner distribution
function, the complex fraction Fourier transformation, the complex wavelet
transformation, and so on \cite{2,3,4,5,9,10}. Among these useful
applications, it is usually necessary to construct the corresponding
asymmetric integration projection operators based on quantum
representations. For example, in order to offer more qualified mother
wavelets for complex wavelet transforms, Fan and Lu introduce the following
asymmetric integration projection operators \cite{3}%
\begin{equation}
U\left( \mu ,\kappa \right) =\frac{1}{\mu }\int \frac{d^{2}\eta }{\pi }|%
\frac{\eta -\kappa }{\mu }\rangle \langle \eta |,  \label{5}
\end{equation}%
where $0\neq \mu \in \mathbb{R}$ and $\kappa \in \mathbb{C}$. Based on Eq. (%
\ref{5}), mother wavelets for complex wavelet transformation can be
considered as a matrix element of the two-mode squeezed displaced operator $%
U\left( \mu ,\kappa \right) $. \ For detailed discussion about the
applications of the entangled state representation for complex wavelet
transformation, please see Ref. \cite{3,5}. In the case of $\kappa =0$, Eq. (%
\ref{5}) is a natural representation of the two-mode squeezed operator $%
U\left( \mu ,0\right) =\exp \left[ \lambda \left( \hat{a}^{\dag }\hat{b}%
^{\dag }-\hat{a}\hat{b}\right) \right] $ with $\mu =e^{\lambda }$ being a
squeezing parameter \cite{6}.

In the previous studies, those constructing asymmetric integration
projection operators based on different quantum representations are usually
quantum unitary operators. For example, for many optical transformations
based on entangled state representations $\left\vert \eta \right\rangle $
and $\left\vert \xi \right\rangle $, those corresponding asymmetric
integration projection operators are most unitary operators \cite{2}.
Different from the past, in this work we will introduce a new kind of the
asymmetric integration projection operators based on the two $\left\vert
\eta \right\rangle $ and $\left\vert \xi \right\rangle $, but these
operators are Hermitian operators. Concretely, we construct the asymmetric
integration projection operators as follows
\begin{equation}
\int \frac{d\eta _{1}d\eta _{2}}{\pi }|\eta _{1}+i\eta _{2}\rangle \langle
\eta _{2}+i\eta _{1}|=?\text{ }  \label{6}
\end{equation}%
or
\begin{equation}
\int \frac{d\xi _{1}d\xi _{2}}{\pi }|\xi _{1}+i\xi _{2}\rangle \langle \xi
_{2}+i\xi _{1}|=?  \label{7}
\end{equation}%
Obviously, we are only swapping the real and imaginary parts of the variable
$\eta $ (or $\xi $) in the bra of the completeness relation of the $%
\left\vert \eta \right\rangle $ (or $\left\vert \xi \right\rangle $), which
leads to that Eqs. (\ref{6}) and (\ref{7}) are no longer a two-mode identity
operator. According to orthonormal relations of the two $\left\vert \eta
\right\rangle $ and $\left\vert \xi \right\rangle $ in Eqs. (\ref{2}) and (%
\ref{4}), one can easily prove that the above asymmetric integration
projection operators are Hermitian operators. On the other hand, it is well
known, in quantum mechanics, for a Hermitian operator, there may exist a
corresponding physical measurement or observable. Enlighten by the previous
work \cite{a1}, we can analytically demonstrate that such Hermitian
operators indeed represent to a physical measurement or observable, which is
exactly the parity measurement combined with a beam splitter in quantum
metrology.

This paper is organized as follows. In Sec. 2, we introduce a kind of
Hermitian operators constructed by entangled state representations. In Sect.
3, we analytically prove that such Hermitian operators represent the parity
measurement combined with a beam splitter. In Sec. 4, we show some
applications of our results. Finally, our conclusions are presented in Sect.
5.

\section{Hermitian operators expressed in entangled state representation}

In quantum mechanics, the IWOP is a useful tool about the transformation
theory of quantum representations and the operations of quantum operators
\cite{1,11}. In this section, we perform these asymmetric integration
expressed in Eqs. (\ref{6}) and (\ref{7}) by IWOP. In order to facilitate
the understanding the IWOP, here we will give a detailed derivation process.
Note that Eq. (\ref{6}) and the normal ordering form of the vacuum
projective operator \cite{1}%
\begin{equation}
|0\rangle _{a}|0\rangle _{b}\,{_{b}\langle 0|}\,{_{a}\langle 0|}=\colon e^{-%
\hat{a}^{\dag }\hat{a}-\hat{b}^{\dag }\hat{b}}\colon ,  \label{8}
\end{equation}%
the left of Eq. (\ref{6}) can be written as

\begin{eqnarray}
&&\int \frac{d\eta _{1}d\eta _{2}}{\pi }|\eta _{1}+i\eta _{2}\rangle \langle
\eta _{2}+i\eta _{1}|  \notag \\
&=&\int \frac{d\eta _{1}d\eta _{2}}{\pi }e^{-\frac{1}{2}\left( \eta
_{1}^{2}+\eta _{2}^{2}\right) +\left( \eta _{1}+i\eta _{2}\right) \hat{a}%
^{\dag }-\left( \eta _{1}-i\eta _{2}\right) \hat{b}^{\dag }+\hat{a}^{\dag }%
\hat{b}^{\dag }}  \notag \\
&&\colon e^{-\hat{a}^{\dag }\hat{a}-\hat{b}^{\dag }\hat{b}}\colon e^{-\frac{1%
}{2}\left( \eta _{1}^{2}+\eta _{2}^{2}\right) +\left( \eta _{2}-i\eta
_{1}\right) \hat{a}-\left( \eta _{2}+i\eta _{1}\right) \hat{b}+\hat{a}\hat{b}%
}.  \label{9}
\end{eqnarray}%
We can see that the right of operator $\colon e^{-\hat{a}^{\dag }\hat{a}-%
\hat{b}^{\dag }\hat{b}}\colon $ are all annihilation operators, while on its
left are all creation operators; therefore the whole integral is in normal
ordering of operators. Then, we can rewrite Eq. (\ref{9}) as the following
form%
\begin{eqnarray}
&&\int \frac{d\eta _{1}d\eta _{2}}{\pi }|\eta _{1}+i\eta _{2}\rangle \langle
\eta _{2}+i\eta _{1}|  \notag \\
&=&\int \frac{d\eta _{1}d\eta _{2}}{\pi }\colon e^{-\left( \eta
_{1}^{2}+\eta _{2}^{2}\right) +\eta _{1}\left( \hat{a}^{\dag }-\hat{b}^{\dag
}-i\hat{a}-i\hat{b}\right) +\eta _{2}\left( i\hat{a}^{\dag }+i\hat{b}^{\dag
}+\hat{a}-\hat{b}\right) }  \notag \\
&&\times e^{\hat{a}^{\dag }\hat{b}^{\dag }+\hat{a}\hat{b}-\hat{a}^{\dag }%
\hat{a}-\hat{b}^{\dag }\hat{b}}\colon  \label{10}
\end{eqnarray}%
Note that the IWOP and the following mathematic integration formula

\begin{equation}
\int_{-\infty }^{\infty }\exp \left[ -\alpha x^{2}+\beta x\right] dx=\sqrt{%
\frac{\pi }{\alpha }}\exp \left( \frac{\beta ^{2}}{4\alpha }\right) ,
\label{11}
\end{equation}%
which holds for {Re}$\left( \alpha \right) >0$, we can directly calculate
the integration in Eq. (\ref{10}) and obtain%
\begin{equation}
\int \frac{d\eta _{1}d\eta _{2}}{\pi }|\eta _{1}+i\eta _{2}\rangle \langle
\eta _{2}+i\eta _{1}|=\colon e^{-i\hat{a}^{\dag }\hat{b}+i\hat{a}\hat{b}%
^{\dag }-\hat{a}^{\dag }\hat{a}-\hat{b}^{\dag }\hat{b}}\colon ,  \label{12}
\end{equation}%
which is a concise expression of operators.

In order to get much more information from Eq. (\ref{12}), we remove the
restricted normal ordering symbol $\colon \colon $, the operators'
non-commutative property manifestly appears. So let us start with Taylor
expansion of $e^{-i\hat{a}^{\dag }\hat{b}}$ and $e^{i\hat{a}\hat{b}^{\dag }}$
in Eq. (\ref{12}), i.e.,%
\begin{eqnarray}
&&\colon e^{-i\hat{a}^{\dag }\hat{b}+i\hat{a}\hat{b}^{\dag }-\hat{a}^{\dag }%
\hat{a}-\hat{b}^{\dag }\hat{b}}\colon   \notag \\
&=&\colon \sum_{m=0,n=0}^{\infty }\frac{\hat{a}^{\dag m}\hat{b}^{m}\left(
-i\right) ^{m}}{m!}e^{-\hat{a}^{\dag }\hat{a}-\hat{b}^{\dag }\hat{b}}\frac{%
\hat{a}^{n}\hat{b}^{\dag n}\left( i\right) ^{n}}{n!}\colon .  \label{13}
\end{eqnarray}%
According to major properties of normally ordered product of operators which
means all the bosonic creation operators $\hat{a}^{\dag }$ ($\hat{b}^{\dag }$%
) are standing on the left of annihilation operators $\hat{a}$ ($\hat{b}$)
in a monomial of $\hat{a}$ ($\hat{b}$) and $\hat{a}^{\dag }$ ($\hat{b}^{\dag
}$)\cite{12}, we further rewrite Eq. (\ref{13}) as%
\begin{eqnarray}
&&\colon e^{-i\hat{a}^{\dag }\hat{b}+i\hat{a}\hat{b}^{\dag }-\hat{a}^{\dag }%
\hat{a}-\hat{b}^{\dag }\hat{b}}\colon   \notag \\
&=&\sum_{m=0,n=0}^{\infty }i{^{\left( n-m\right) }}\frac{\hat{a}^{\dag m}%
\hat{b}^{\dag n}}{\sqrt{m!n!}}\colon e^{-\hat{a}^{\dag }\hat{a}-\hat{b}%
^{\dag }\hat{b}}\colon \frac{\hat{a}^{n}\hat{b}^{m}}{\sqrt{m!n!}}.
\label{14}
\end{eqnarray}%
And then, applying Eq. (\ref{8}), we finally have%
\begin{eqnarray}
&&\colon e^{-i\hat{a}^{\dag }\hat{b}+i\hat{a}\hat{b}^{\dag }-\hat{a}^{\dag }%
\hat{a}-\hat{b}^{\dag }\hat{b}}\colon   \notag \\
&=&\sum_{m=0,n=0}^{\infty }i{^{\left( n-m\right) }}|m\rangle _{a}|n\rangle
_{b}\,{_{b}\langle m|}\,{_{a}\langle n|.}  \label{15}
\end{eqnarray}%
In Eq. (\ref{15}), we have used the expression of the two-mode Fock state $%
\left\vert m\right\rangle _{a}\left\vert n\right\rangle _{b}=\frac{\hat{a}%
^{\dag m}\hat{b}^{\dag n}}{\sqrt{m!n!}}\left\vert 0\right\rangle
_{a}\left\vert 0\right\rangle _{b}$. Comparing with the completeness
relation of the two-mode Fock state, we can see that $m$ and $n$ in two bras
are exchanged and an additional item $i{^{\left( n-m\right) }}$ is
introduced. Here, we must point out that Eq. (\ref{6}) is an asymmetric
integration projection operator with continuous variables, while Eq. (\ref%
{15}) is an asymmetric sum projection operator with discrete variables.
However, they represent the same Hermitian operator.

Now, we turn to investigate the asymmetric integration projection operator
based on entangled state $\left\vert \xi \right\rangle $. By the similar
way, we can complete the integration of Eq. (\ref{7}) and obtain
\begin{equation}
\int \frac{d\xi _{1}d\xi _{2}}{\pi }|\xi _{1}+i\xi _{2}\rangle \langle \xi
_{2}+i\xi _{1}|=\colon e^{i\hat{a}^{\dag }\hat{b}-i\hat{a}\hat{b}^{\dag }-%
\hat{a}^{\dag }\hat{a}-\hat{b}^{\dag }\hat{b}}\colon   \label{16}
\end{equation}%
which is some different from Eq. (\ref{12}). But in essence they are a kind
of Hermitian operators, which will be proved in the following. Of course, we
can also cast such normal ordering operator $\colon \exp \left[ i\hat{a}%
^{\dag }\hat{b}-i\hat{a}\hat{b}^{\dag }-\hat{a}^{\dag }\hat{a}-\hat{b}^{\dag
}\hat{b}\right] \colon $ into the Fock representation, i.e.,
\begin{equation}
\colon e^{i\hat{a}^{\dag }\hat{b}-i\hat{a}\hat{b}^{\dag }-\hat{a}^{\dag }%
\hat{a}-\hat{b}^{\dag }\hat{b}}\colon =\sum_{m=0,n=0}^{\infty }i{^{\left(
m-n\right) }}|m\rangle _{a}|n\rangle _{b}\,{_{b}\langle m|}\,{_{a}\langle n|}
\label{17}
\end{equation}%
Obviously, Eq. (\ref{17}) is not a two-mode identity operator. Equations (%
\ref{16}) and (\ref{17}) also represents the same Hermitian operator. In the
next, we will analytically prove that such Hermitian operators in Eqs. (\ref%
{12}) and (\ref{16}) correspond to a parity measurement of a two-mode
quantum state after passing through a beam splitter.

\section{Physical correspondence of a kind of asymmetric integration
projection operator}

In the quantum metrology, a Mach-Zehnder interferometer is a useful tool to
estimate a slight variation of phase shift originating from different
physical processes. A typical Mach-Zehnder interferometer is composed of two
beam splitters and the phase shift between the two arms to be estimated as
shown in Fig. 1. In order to extract the phase shift, one need to choose a
special measurement, such as intensity measurement, homodyne measurement,
and parity measurement \cite{13}. Here, we consider parity measurement on
one output of the interferometer.

\begin{figure}[tbph]
\centering
\includegraphics[width=8cm]{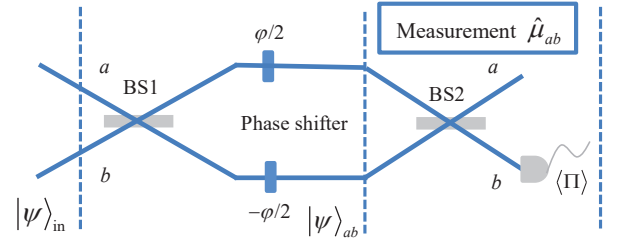}
\caption{(color online) Sketch of parity measurement of a two-mode quantum
state passing through a beam splitter.}
\end{figure}

Note that performing parity measurement onto one output of the
interferometer is equivalent to compute the expectation value of the parity
operator $\hat{\Pi}=\exp \left[ i\pi \hat{b}^{\dag }\hat{b}\right] $ in the
output state. From the perspective of metrology, all the devices after the
phase shift are regarded as a measurement as shown in Fig. 1. Therefore, the
parity measurement combined with a beam splitter is equivalent to the
implementation of projective measurement to the state ${|\psi \rangle _{ab}}$
before the second beam splitter, i.e.,%
\begin{eqnarray}
\left\langle \hat{\Pi}\right\rangle  &=&{_{ab}\langle \psi |}\hat{U}%
_{BS}^{\dag }\left[ \hat{I}_{A}\otimes \exp \left[ i\pi \hat{b}^{\dag }\hat{b%
}\right] \right] \hat{U}_{BS}\,{|\psi \rangle _{ab}}  \notag \\
&=&{_{ab}\langle \psi |\hat{\mu}}_{ab}{|\psi \rangle _{ab}},  \label{18}
\end{eqnarray}%
where $\hat{I}_{A}$ represents a unity operator of $a$ mode of the two-mode
quantum system, $\hat{U}_{BS}$ is the unitary operator for the beam
splliter, and the projective operator ${\hat{\mu}}_{ab}$ is

\begin{equation}
{\hat{\mu}}_{ab}=\hat{U}_{BS}^{\dag }\left[ \hat{I}_{A}\otimes \exp \left[
i\pi \hat{b}^{\dag }\hat{b}\right] \right] \hat{U}_{BS}.  \label{19}
\end{equation}%
It should be pointed out that the use of ${\hat{\mu}}_{ab}$ here highlights
the fact that parity measurement combined with a beam splitter provides a
measurement scheme including all of the phase-carrying off-diagonal terms in
the two-mode density matrix \cite{14,15}. In quantum optics, a general beam
splitter which can be described by the following unitary operator \cite%
{a2,a3}

\begin{equation}
\hat{U}_{BS}=\exp \left[ \frac{\theta }{2}\left( \hat{a}^{\dagger }\hat{b}%
e^{i\phi }-\hat{a}\hat{b}^{\dag }e^{-i\phi }\right) \right] ,  \label{20}
\end{equation}%
with the following transformation relations%
\begin{eqnarray}
\hat{U}_{BS}^{\dag }\hat{a}\hat{U}_{BS} &=&\hat{a}\cos \frac{\theta }{2}+%
\hat{b}e^{i\phi }\sin \frac{\theta }{2},  \notag \\
\hat{U}_{BS}^{\dag }\hat{b}\hat{U}_{BS} &=&\hat{b}\cos \frac{\theta }{2}-%
\hat{a}e^{-i\phi }\sin \frac{\theta }{2}.  \label{21}
\end{eqnarray}%
For a 50:50 beam splitter with $\phi =0$, the projective operator ${\hat{\mu}%
}_{ab}$ has been casted into the Fock state representation \cite{15,16}%
\begin{equation}
{\hat{\mu}}_{ab}|_{\theta =\pi /2,\phi =0}=\sum\limits_{m,n}^{\infty
}\left\vert m\right\rangle _{a}\left\vert n\right\rangle _{b}\left.
_{b}\left\langle m\right\vert \right. \left. _{a}\left\langle n\right\vert
\right. .  \label{r3}
\end{equation}%
Recently, we have proved analytically that Eq. (\ref{r3}) is indeed valid
\cite{a1}. In addition, we also demonstrate that such projective operator can
be rewritten in the coherent representation, i.e.,%
\begin{equation}
{\hat{\mu}}_{ab}|_{\theta =\pi /2,\phi =0}=\frac{d^{2}\alpha d^{2}\beta }{%
\pi ^{2}}\left\vert \alpha \right\rangle _{a}\left\vert \beta \right\rangle
_{b}\left. _{b}\left\langle \alpha \right\vert \right. \left.
_{a}\left\langle \beta \right\vert \right. ,  \label{r4}
\end{equation}%
which is another kind of asymmetric integration projection operator. In Eq. (%
\ref{r4}), the variables $\alpha $ and $\beta $ in two bras are exchanged
compared with the case of the two kets. Of course, the product state $%
\left\vert \alpha \right\rangle _{a}\left\vert \beta \right\rangle _{b}\,$is
not a entangled state. In this work, the asymmetric integration projection
operators are constructed by entangled quantum representations as shown in
Eqs. (\ref{6}) and (\ref{7}). Here, we are only swapping the real and
imaginary parts of the variable $\eta $ (or $\xi $) in the bras of the
completeness relations of $\left\vert \eta \right\rangle $ and $\left\vert
\xi \right\rangle $. Although, in form at least, the Eq. (\ref{r4}) is
somewhat different from Eqs. (\ref{6}) and (\ref{7}), such Hermitian
operators can be proved in the following that they indeed represent to the
same physical measurement or observable.

Now, for our purpose, we will demonstrate that these asymmetric projection
operators Eqs. (\ref{6}) and (\ref{7}) also describe the parity measurement
with a balanced beam splitter as shown in Fig. 1. Based on the coherent
state representation, the parity operator can be expressed as \cite{17}
\begin{equation}
\left( -1\right) ^{\hat{b}^{\dag }\hat{b}}=\exp \left[ i\pi \hat{b}^{\dag }%
\hat{b}\right] =\int \frac{d^{2}\beta }{\pi }|\beta \rangle _{b}{_{b}\langle
-\beta |}.  \label{22}
\end{equation}%
Substituting $\hat{I}_{A}=\int d^{2}|\alpha \rangle _{a}\,{_{a}\langle
\alpha |}/\pi $ in the coherent state representation and Eqs. (\ref{21}) and
(\ref{22}) into Eq. (\ref{19}), we perform the integration and can obtain
the normal ordering form of the projective operator ${\hat{\mu}}_{ab}$ of the
parity measurement
\begin{equation}
{\hat{\mu}}_{ab}=\colon e^{\hat{a}^{\dagger }\hat{a}\cos \theta -\hat{b}%
^{\dag }\hat{b}\cos \theta +\hat{a}^{\dagger }\hat{b}e^{i\phi }\sin \theta +%
\hat{a}\hat{b}^{\dag }e^{-i\phi }\sin \theta -\hat{a}^{\dag }\hat{a}-\hat{b}%
^{\dag }\hat{b}}\colon ,  \label{23}
\end{equation}%
where we have used the integral formula \cite{18}
\begin{equation}
\int \frac{d^{2}z}{\pi }\exp \left( \zeta \left\vert z\right\vert ^{2}+\xi
z+\eta z^{\ast }\right) =-\frac{1}{\zeta }\exp \left[ -\frac{\xi \eta }{%
\zeta }\right] ,  \label{24}
\end{equation}%
whose convergent condition is Re$\left( \zeta \right) <0$. Particularly. we
consider the case of $\theta =\pi /2$, i.e., for the balanced beam splitter,
and obtain
\begin{equation}
{\hat{\mu}}_{ab}|_{\theta =\pi /2}=\colon \exp \left[ \hat{a}^{\dag }\hat{b}{%
e^{i\phi }}+\hat{a}\hat{b}^{\dag }{e^{-i\phi }}-\hat{a}^{\dag }\hat{a}-\hat{b%
}^{\dag }\hat{b}\right] \colon ,  \label{25}
\end{equation}%
or
\begin{equation}
{\hat{\mu}}_{ab}|_{\theta =\pi /2}=\sum_{m=0,n=0}^{\infty }e^{i\left(
m-n\right) \phi }|m\rangle _{a}|n\rangle _{b}\,{_{b}\langle m|}\,{%
_{a}\langle n|},  \label{26}
\end{equation}%
Eq. (\ref{26}) is the Fock representation of the parity measurement combined
with a balanced beam splitter. Here, our result Eq. (\ref{26}) is somewhat
different from Eq. (\ref{r3}) appeared in Refs. \cite{15,16}. From Eq. (\ref%
{26}), we can see that $m$ and $n$ in two bras are exchanged and an
additional item $e^{i\left( m-n\right) \phi }$ is introduced. Naturally, in
the case of $\phi =0$, Eq. (\ref{26}) just is that result in Refs. \cite%
{15,16}.

Now, comparing Eq. (\ref{25}) with Eq. (\ref{12}), we easily obtain the
following relation in the case of $\phi =-\pi /2$,

\begin{equation}
\int \frac{d\eta _{1}d\eta _{2}}{\pi }|\eta _{1}+i\eta _{2}\rangle \langle
\eta _{2}+i\eta _{1}|={\hat{\mu}}_{ab}|_{\theta =\pi /2,\phi =-\pi /2}.
\label{27}
\end{equation}%
Similarly, in the case of $\phi =\pi /2$, comparing Eq. (\ref{25}) with Eq. (%
\ref{16}), we have
\begin{equation}
\int \frac{d\xi _{1}d\xi _{2}}{\pi }|\xi _{1}+i\xi _{2}\rangle \langle \xi
_{2}+i\xi _{1}|={\hat{\mu}}_{ab}|_{\theta =\pi /2,\phi =\pi /2}.  \label{28}
\end{equation}%
Equations (\ref{27}) and \ref{28} show a new relation between a Hermitian
operator and the entangled state representations, which is one important
result\ in this work. So far, we have proved analytically that those
asymmetric integration projection operators constructed by the two entangled
states $\left\vert \eta \right\rangle $ and $\left\vert \xi \right\rangle $
indeed represent a physical measurement. That is to say, such asymmetric
projection operators also correspond to a parity measurement combined with a
beam splitter. Therefore, for the same measurement operators in quantum
mechanics, one can cast it into different quantum representations according
to the needs of the specific problems. These results not only can improve
the understanding of the relations between the Hermitian operators and
quantum representations, but also can bring us some convenience in dealing
with the calculation related to quantum operators.

\section{Some applications the projective operator of parity measurement}

In quantum optical metrology, besides usual intensity measurement and
homodyne measurement, parity measurement has been shown to be a universal
measurement scheme in phase estimation, and has been demonstrated to be the
optimal measurement strategy for many input states. In order to improve the
phase sensitivity, it has been proved that, with parity measurement, the
Heisenberg limit can be saturated with a lossless Mach-Zehnder
interferometer by using nonclassical states as the inputs, such as squeezed
states \cite{19,20,21}, NOON states \cite{22}, and entangled coherent states
\cite{23}. As applications of our results shown in Eq. (\ref{27}) (or Eq. (%
\ref{28})), in what follows we consider two specific states that have been
frequently investigated in quantum optical metrology.

Case 1.--- Let us first consider the input state of a Mach-Zehnder
interferometer to be an NOON state \cite{22}:%
\begin{equation}
\left\vert \Psi \right\rangle _{\text{NOON}}=\frac{1}{\sqrt{2}}\left(
\left\vert N\right\rangle _{a}\left\vert 0\right\rangle _{b}+\left\vert
0\right\rangle _{a}\left\vert N\right\rangle _{b}\right) .  \label{29}
\end{equation}%
Upon leaving the phase channel and before the second beam splitter as shown
in Fig. 1, the state arrives at%
\begin{eqnarray}
\left\vert \psi \right\rangle _{ab} &=&\hat{U}\left( \varphi \right)
\left\vert \Psi \right\rangle _{\text{NOON}}  \notag \\
&=&\frac{1}{\sqrt{2}}\left( e^{i\varphi N/2}\left\vert N\right\rangle
_{a}\left\vert 0\right\rangle _{b}+e^{-i\varphi N/2}\left\vert
0\right\rangle _{a}\left\vert N\right\rangle _{b}\right) .  \label{30}
\end{eqnarray}%
where $\hat{U}\left( \varphi \right) =\exp \left[ i\varphi \left( \hat{a}%
^{\dag }\hat{a}-\hat{b}^{\dag }\hat{b}\right) /2\right] $ denotes the two
phase shifters, the angle $\varphi $ is the phase shift between the two arms
of a Mach-Zehnder interferometer to be estimated. According to Eqs. (\ref{18}%
) and (\ref{12}) (or (\ref{27})), the expectation value of the parity
measurement scheme $\left\langle \hat{\Pi}\right\rangle $ can be expressed by%
\begin{equation}
\left\langle \hat{\Pi}\right\rangle =\int \frac{d\eta _{1}d\eta _{2}}{\pi }%
_{ab}\left\langle \psi \right\vert \left. \eta _{1}+i\eta _{2}\right\rangle
\left\langle \eta _{2}+i\eta _{1}\right. \left\vert \psi \right\rangle _{ab}.
\label{31}
\end{equation}%
Note that the entangled state $|\eta \rangle $ can be expanded in terms of
two-mode Fock states as \cite{1}
\begin{equation}
|\eta =\eta _{1}+i\eta _{2}\rangle =e^{-\left\vert \eta \right\vert
^{2}/2}\sum_{m,n=0}^{\infty }H_{m,n}\left( \eta ,\eta ^{\ast }\right) \frac{%
\left( -1\right) ^{n}}{\sqrt{m!n!}}\left\vert m\right\rangle _{a}\left\vert
n\right\rangle _{b},  \label{32}
\end{equation}%
where $H_{m,n}\left( \eta ,\eta ^{\ast }\right) $ is the two-variable
Hermite polynomial
\begin{equation}
H_{m,n}\left( \xi ,\eta \right) =\frac{\partial ^{m+n}}{\partial
t^{m}\partial t^{\prime n}}\exp [-tt^{\prime }+t\xi +t^{\prime }\eta
]_{t,t^{\prime }=0}.  \label{33}
\end{equation}%
Thus, substituting Eqs. (\ref{32}) and (\ref{33}) into Eq. (\ref{31}) and
noting that the mathematical integration formula shown in Eq. (\ref{11}),
after some simple calculation we have%
\begin{equation}
\left\langle \hat{\Pi}\right\rangle =\frac{i^{N}}{2}\left[ \exp \left[
iN\varphi \right] +\left( -1\right) ^{N}\exp \left[ -iN\varphi \right] %
\right] ,  \label{34}
\end{equation}%
which is just that results in Ref. \cite{22}.

Case 2.--- We then consider a coherent state combined with a squeezed vacuum
state, i.e., $\left\vert z\right\rangle _{a}\otimes \left\vert
r\right\rangle _{b}$. For the convenience of the later calculation, we
rewrite the squeezed vacuum state $\left\vert r\right\rangle $ in the basis
of the coherent state as follows%
\begin{equation}
\left\vert r\right\rangle _{b}=\mathrm{sech}^{1/2}r\int \frac{d^{2}\alpha }{%
\pi }e^{-\frac{\left\vert \alpha \right\vert ^{2}}{2}-\frac{\tanh r}{2}%
\alpha ^{\ast 2}}\left\vert \alpha \right\rangle _{b},  \label{35}
\end{equation}%
where $\left\vert \alpha \right\rangle =\exp \left[ -\left\vert \alpha
\right\vert ^{2}/2+\alpha b^{\dag }\right] \left\vert 0\right\rangle $ is a
coherent state. When the product state $\left\vert \psi \right\rangle _{%
\text{in}}=\left\vert z\right\rangle _{a}\otimes \left\vert \psi
_{r}\right\rangle _{b}$ is injected into a Mach-Zehnder interferometer, upon
leaving the phase channel, the state evolves as%
\begin{eqnarray}
\left\vert \psi \right\rangle _{ab} &=&\hat{U}\left( \varphi \right) \hat{U}%
_{BS1}\left\vert z\right\rangle _{a}\left\vert r\right\rangle _{b}  \notag \\
&=&\mathrm{sech}^{1/2}r\int \frac{d^{2}\alpha }{\pi }e^{-\frac{\left\vert
\alpha \right\vert ^{2}}{2}-\frac{\tanh r}{2}\alpha ^{\ast 2}}  \notag \\
&&\times \left\vert \frac{\left( z+i\alpha \right) e^{i\varphi /2}}{\sqrt{2}}%
\right\rangle _{a}\left\vert \frac{\left( iz+\alpha \right) e^{-i\varphi /2}%
}{\sqrt{2}}\right\rangle _{b}\text{ },  \label{36}
\end{eqnarray}%
where $\hat{U}_{BS1}=\exp \left[ i\pi \left( \hat{a}^{\dagger }\hat{b}+\hat{a%
}\hat{b}^{\dag }\right) /4\right] $ describes the first beam splitter.
Similarly, applying Eq. (\ref{12}) or (\ref{27}) we have

\begin{eqnarray}
\left\langle \hat{\Pi}\right\rangle  &=&\int \frac{d\eta _{1}d\eta _{2}}{\pi
}_{ab}\left\langle \psi \right\vert \left. \eta _{1}+i\eta _{2}\right\rangle
\left\langle \eta _{2}+i\eta _{1}\right. \left\vert \psi \right\rangle _{ab}
\notag \\
&=&\left. _{ab}\left\langle \psi \right\vert \right. \colon e^{-i\hat{a}%
^{\dag }\hat{b}+i\hat{a}\hat{b}^{\dag }-\hat{a}^{\dag }\hat{a}-\hat{b}^{\dag
}\hat{b}}\colon \left\vert \psi \right\rangle _{ab}.  \label{37}
\end{eqnarray}%
Therefore, substituting Eq. (\ref{36}) into Eq. (\ref{37}), and noting that
the following integral formula \cite{18}
\begin{equation}
\int \frac{d^{2}z}{\pi }e^{\zeta \left\vert z\right\vert ^{2}+\xi z+\eta
z^{\ast }+fz^{2}+gz^{\ast 2}}=\frac{1}{\sqrt{\zeta ^{2}-4fg}}e^{\frac{-\zeta
\xi \eta +\xi ^{2}g+\eta ^{2}f}{\zeta ^{2}-4fg}},  \label{38}
\end{equation}%
whose convergent condition is Re$\left( \xi \pm f\pm g\right) <0$ and$\ $Re$%
\left( \frac{\zeta ^{2}-4fg}{\xi \pm f\pm g}\right) <0$, by performing this
integral we have%
\begin{equation}
\left\langle \Pi _{b}\left( \varphi \right) \right\rangle _{0}=\frac{e^{%
\frac{2\left( \cos \varphi -1-\sinh ^{2}r\sin ^{2}\varphi \right) \left\vert
z\right\vert ^{2}-\sinh 2r\sin ^{2}\varphi \mathrm{Re}\left( z^{2}\right) }{%
2\left( 1+\sinh ^{2}r\sin ^{2}\varphi \right) }}}{\sqrt{1+\sinh ^{2}r\sin
^{2}\varphi }}.  \label{39}
\end{equation}%
which is the corresponding expectation value of the parity operator when a
combination of coherent state and squeezed vacuum state is considered as
input states of a Mach-Zehnder interferometer \cite{21}. For other quantum
states as the interferometer state, in general we can also obtain the
expectation value of the parity measurement scheme by the above method.
Therefore, our results offer a new method to investigate the parity-based
phase estimation scheme with calculations of its signal.

\section{Conclusion}

In summary, we construct a new kind of asymmetric integration projection
operators based on the entangled state representation, which are the
Hermitian. Then, we analytically demonstrate what physical observable does
such new asymmetric projection operators represent. Our results show
explicitly that such Hermitian operator describes the effect of the parity
measurement with a balanced beam splitter. Finally, as applications of our
results, we regain the expected signal of the parity-based phase estimation
scheme with some specific states that have been frequently investigated in
quantum optical metrology. In a word, for some measurement operators in
quantum mechanics, they may be casted into different quantum
representations, which can improve the understanding of the relations
between the Hermitian operators and quantum representations.

\section*{Acknowledgments}

This work is supported by the National Natural Science Foundation of China
(No. 11905160 and No. 11404040) and sponsored by Qing Lan Project of the
Higher Educations of Jiangsu Province of China.

\end{document}